\documentclass[aps,prl,twocolumn,showpacs]{revtex4-1}
\pdfoutput=1
\usepackage{amsmath}    
\usepackage{graphicx}   
\usepackage{color}      
\usepackage{hyperref}   

\usepackage{amsmath}

\graphicspath{{./Figures/}}
\hypersetup{pdfborder = {0 0 0},colorlinks=true}

\renewcommand{\H}{{\mathcal H}}
\newcommand{\woone}{_{\backslash 1}}

\newcommand{\J}{\boldsymbol{J}} 
\newcommand{\s}{\boldsymbol{s}} 
\newcommand{\Zstar}{Z^{\star}} 
\newcommand{\h}{k} 
\newcommand{\llangle}{\langle\!\langle}
\newcommand{\rrangle}{\rangle\!\rangle}
 \newcommand{\para}[1]{{\bf #1.}}

\begin{document}
\title{Statistical mechanics of the inverse Ising problem and the optimal objective function}

\author{Johannes Berg}
\email{berg@thp.uni-koeln.de}
\affiliation{Institute for Theoretical Physics, University of Cologne,
  Z\"ulpicher Stra{\ss}e 77, 50937 Cologne, Germany }

\begin{abstract}
The inverse Ising problem seeks to reconstruct the parameters of an Ising Hamiltonian on the basis of spin configurations sampled from the Boltzmann measure. 
Over the last decade, many applications of the inverse Ising problem have arisen, driven by the advent of large-scale data across different scientific disciplines. 
Recently, strategies to solve the inverse Ising problem based on convex optimisation have proven to be very successful. These approaches maximise particular objective functions with respect to the model parameters. Examples are the pseudolikelihood method and interaction screening. In this paper, we establish a link between approaches to the inverse Ising problem based on convex optimisation and the statistical physics of disordered systems. 
We characterise the performance of an arbitrary objective function and calculate the objective function which optimally reconstructs the model parameters. We evaluate the optimal objective function within a replica-symmetric ansatz and compare the results of the optimal objective function with other reconstruction methods. Apart from giving a theoretical underpinning to solving the inverse Ising problem by convex optimisation, the optimal objective function outperforms state-of-the-art methods, albeit by a small margin. 
 \end{abstract}




\pacs{02.30.Zz,02.50.Tt,89.75.-k, 75.50.Lk}

\maketitle

The advent of large-scale data across different scientific disciplines, especially biology, has inspired many applications of the inverse Ising problem. 
Over the last decade, the inverse Ising problem has been used to analyze neural firing patterns~\cite{Schneidman2006a} and gene expression data~\cite{lezon2006}, to infer biological fitness landscapes~\cite{mora2010a,shekhar2013a}, and to analyze financial data~\cite{bury2013a}. A variant of the inverse Ising model with more than two states for each spin has been used to determine the three-dimensional structure of proteins~
\cite{Weigt2009a,ekeberg2013,ovchinnikov2017,coccoetal2017}. This versatility is not surprising: the inverse Ising problem arises naturally when one wants to learn the interactions between discrete random variables describing an equilibrium system. For a review, see~\cite{nguyenzecchinaberg2017}.

Conceptually, the inference of parameters of an Ising model from data is a simple matter: Consider an Ising model 
with $N$ binary spin variables, $s_i = \pm1, i=1,\ldots,N$. Pairwise interactions between the spins lead to the well-known Ising Hamiltonian
\begin{equation}
\H= -\sum_{i<j} \frac{J_{ij}^{\star}}{\sqrt{N}} s_i s_j , 
\end{equation}
where  $\frac{J_{ij}^{\star}}{\sqrt{N}}$ quantifies the coupling strength between a pair of spins, which we seek to infer. We have inserted a constant $1/\sqrt{N}$ for later convenience, magnetic fields can also be added without difficulty. 
$M$ spin configurations (samples) $\s^1,\ldots,\s^M$ are drawn independently from the Boltzmann distribution
\begin{equation}
\label{eq:boltzmann} 
P_B(\s|\J^{\star})=\frac{1}{\Zstar({\bf J^{\star}})} \exp\left\{\sum_{i<j} \frac{J_{ij}^{\star}}{\sqrt{N}} s_i s_j \right\} \ ,
\end{equation}
and the task is to find the couplings which produce these spin configurations. This can be done by maximizing the so-called log-likelihood 
\begin{equation}
\label{eq:likelihood} 
\sum_{\mu=1}^M \ln P_B(\s^{\mu}|\J) = \sum_{i<j} \frac{J_{ij}}{\sqrt{N}} \sum_{\mu=1}^M s_i^{\mu} s_j^{\mu} - M \ln Z^{\star}(\J) \ ,
\end{equation}
with respect to the couplings, which yields
the maximum-likelihood estimate of the couplings. Alternatively,  
Bayes theorem specifies a probability distribution over the reconstructed couplings
\begin{equation}
\label{eq:posterior} 
P(\J|\s^1,\ldots,\s^M)=\frac{\prod_{\mu=1}^M P_B(\s^{\mu}|\J) P(\J)}{P(\s^1,\ldots,\s^M)} 
\end{equation}
called the posterior probability. One can reconstruct the couplings by maximizing this posterior probability with respect to the couplings, or by computing their expected value under the posterior. In the limit of a large number of samples $M/N \to \infty$, maximizing the Bayesian posterior~\eqref{eq:posterior} yields the same couplings as maximizing the log-likelihood
~\eqref{eq:likelihood}.

In practice, however, the computation of either the likelihood or the Bayesian posterior is a hard task: the Boltzmann distribution~\eqref{eq:boltzmann} contains the partition function, whose computations requires a number of steps which scales exponentially with the system size.
A large number of approaches to likelihood maximisation have been made using the tools of statistical physics, including Monte Carlo methods for small systems~\cite{broderick2007faster}, 
the mean field approximation~\cite{kappen1998b}, a small-correlation expansion~\cite{Sessak2009a}, and others. However, 
one of the most successful methods to solve the inverse Ising problem sidesteps the computation of the likelihood altogether. It originates from statistics and is called pseudolikelihood~\cite{Besag1974a,Ravikumar2010a,Aurell2012b}. 
Pseudolikelihood reconstruction proceeds by maximizing 
\begin{align}
\label{eq:pl_rho}
\sum_{\mu}&  s_i^{\mu} \sum_{j\neq i} \frac{ J_{ij}}{\sqrt{N}} s_j^{\mu}- \ln\left( 2 \cosh \left(\sum_{j\neq i}\frac{ J_{ij}}{\sqrt{N}} s_j^{\mu}\right) \right) \nonumber\\
&= \sum_{\mu} \h_i^{\mu}-\ln (2 \cosh \h_i^{\mu})=\sum_{\mu} \rho(\h_i^{\mu})
\end{align}
with respect to the couplings, or rather, with respect to a particular row $J_{i\bullet}$ of the matrix of couplings. 
In~\eqref{eq:pl_rho}, we have introduced a shorthand describing spins coupled their effective local field $\h_i^{\mu}\equiv s_i^{\mu} \sum_{j \neq i} \frac{ J_{ij}}{\sqrt{N}} s_j^{\mu}$, as well as the pseudolikelihood objective function $\rho(\h)=\h-\ln (2 \cosh \h)$. This method can be interpreted as using a paramagnetic model to describe the statistics of one particular spin $s_i$ in an effective local field, which depends on the couplings between spins. Pseudolikelihood reconstruction has a number of attractive features: the couplings can be determined row-by-row using a convex optimisation algorithm, and the reconstruction becomes exact in the limit $M/N \to \infty$, even 
at low temperatures where many other methods fail~\cite{Aurell2012b}. 
In this way, the couplings $J_{ij}$ and $J_{ji}$ are inferred independently; a symmetric coupling matrix 
can be obtained by considering $(J_{ij} + J_{ji})/2$, although alternatives are possible~\cite{Aurell2012b}.

Recently, a different function has been proposed as an objective function, $\rho(\h)=e^{-\h}$: 
\begin{equation}
\label{eq:interaction_screening}
 \sum_{\mu} e^{-s_i^{\mu} \sum_j \frac{ J_{ij}}{\sqrt{N}} s_j^{\mu}} = \sum_{\mu} e^{-\h_i^{\mu}}
\end{equation}
is to be minimised over the row $J_{i\bullet}$ of the matrix of couplings~\cite{vuffray2016interaction,lokhov2016optimal}. This reconstruction method, termed interaction screening, outperforms pseudolikelihood when the underlying coupling matrix is sparse, and comes close to saturating bounds 
on reconstruction set by information theory~\cite{santhanam2012information}. 

Given these two objective functions, one can ask if there is an objective function $\rho_{\text{opt}}(\h)$, which reconstructs the parameters of the Ising model optimally, that is, minimises the difference between the reconstructed and underlying couplings over all functions $\rho(\h)$ that one might use. In this paper, we build a statistical mechanics of the inverse Ising problem based on the family of objective functions $\rho(\h)$. 
This theory 
tells us how well a certain objective function reconstructs the underlying couplings. It can also be used to derive the objective function which performs best. 
The theory applies to typical realisations of the underlying couplings, which in the thermodynamic limit $N\to \infty$ are realised in nearly all instances of the couplings drawn from a particular distribution. For simplicity, we restrict ourselves to coupling matrices whose entries are drawn independently from a Gaussian distribution.

\para{The partition function for the inverse problem} We start by considering an arbitrary (convex) objective function $\rho(\h)$. 
A regularizing term will be added below.
For the first row of the coupling matrix $\textbf{J}_1=\{J_{1j}\}$ (and equivalently for all other rows) we obtain the minimum of the objective function 
\begin{align}
\label{eq:partitionJopt}
\text{min}_{\{J_{1j}\}} &  \left[ \sum_{\mu=1}^M \rho( \frac{s_1^{\mu}}{\sqrt{N}} \sum_{j\neq 1} J_{1j} s_j^{\mu})\right]\\
&=-\lim_{\beta \to \infty} \partial_{\beta}\ln \int d \textbf{J}_1 e^{-\beta \sum_{\mu=1}^M \rho(\h_1^{\mu})} \nonumber\\
&=-\lim_{\beta \to \infty} \partial_{\beta} 
\ln Z ({\bf{s}}^1,{\bf{s}}^2,\ldots,{\bf{s}}^M) \nonumber \ ,
\end{align}
from the partition function for the inverse problem we define as $Z({\bf{s}}^1,\ldots,{\bf{s}}^M)=\int d {\textbf{J}}_1 e^{-\beta \sum_{\mu=1}^M \rho(\h_1^{\mu})}$. In this partition function, the 
exponential function plays the role of a Boltzmann weight, from which the limit $\beta \to \infty$ selects the ground state thus minimizing $ \sum_{\mu=1}^M \rho(\h_1^{\mu})$. 
The $M$ spins samples ${\bf{s}}^1,\ldots,{\bf{s}}^M$ are taken independently from the Boltzmann distribution~\eqref{eq:boltzmann}. The spin samples can be considered as quenched disorder and  remain fixed while the minimum over $\textbf{J}_1$ is sought.
Conversely, the entries of the reconstructed matrix of couplings
act as phase space variables. The logarithm of the partition function averaged over the disorder, the so-called quenched average, is 
\begin{align}
\label{eq:big_partition0}
\llangle \ln Z&(\textbf{s}^1,\ldots,\textbf{s}^M) \rrangle= \prod_{i<j} \int \frac{dJ_{ij}^{\star}}{\sqrt{2 \pi q^{\star}}} \exp\left\{ - \frac{1}{2q^{\star}} \sum_{i<j} J_{ij}^{*2} \right\}\nonumber\\
&\prod_{\mu}\left( \frac{1}{\Zstar({\bf J^{\star}})} \sum_{\textbf{s}^{\mu}}\right) \exp\left\{ \sum_{i<j,\mu} \frac{ J_{ij}^{\star}}{\sqrt{N}} s_i^{\mu} s_j^{\mu} \right\}\\
&\ln  \left(\int d\textbf{J}_1 \exp\left\{ -\beta \sum_{\mu} \rho( \frac{s_1^{\mu}}{\sqrt{N}} \sum_{j \neq 1} J_{1j} s_j^{\mu} ) \right\} \right)\nonumber\ ,
\end{align}
where the double pointed brackets indicate the average both over the underlying couplings and samples. $q^{\star}$ denotes the variance of the underlying couplings. This averaged partition function describes the parameter inference for a typical 
realisation of the couplings and the spin configurations the reconstruction is based on. Partition functions of this type, where the couplings play the role of phase space variables, have been studied widely in the context of statistical learning
~\cite{watkin1993a,engel2001a}. Bachschmid-Romano and Opper used such a partition function specifically to characterize the reconstruction of the Ising model with asymmetric couplings~\cite{bachschmid2015learning} and have 
recently extended their analysis to the case of symmetric couplings considered here~\cite{bachschmid2017statistical} (see conclusion). A related statistical problem, which can also be addressed using a similar 
partition function, is regression~\cite{advani2016statistical}.

To evaluate the partition function~\eqref{eq:big_partition0}, we use the replica-trick in two different places: to represent the logarithm of the partition function $Z({\bf{s}}^1,\ldots,{\bf{s}}^M)$ and to compute one over the partition function  $\Zstar({\bf J^{\star}})$ in the Boltzmann measure. We obtain the free energy
\begin{align}
\label{eq:free_energy}
- f &= \lim_{\beta \to \infty} \frac{1}{\beta}\llangle \ln Z(\textbf{s}^1,\ldots,\textbf{s}^M) \rrangle\\
&=\mbox{extr}_{q,v,R} \left[\frac{q-R^2/q^{\star}}{2v}  - \alpha \int Dt  {\mathcal M}_v[\rho](R+\sqrt{q}t) \right] \ , \nonumber
\end{align}
with $Dt=\frac{dt}{\sqrt{2 \pi}} e^{-t^2/2}$ and $\alpha=M/N$. This free energy is evaluated by extremizing over the order-parameters $q, R$ and $v$. 
The result is based on a small couplings (low $q^{\star}$) expansion summed to 
infinite order, a replica-symmetric ansatz, the low-temperature limit $\beta \to \infty$, and the thermodynamic limit $N \to \infty$.  Details can be found in the Methods section. 
\begin{equation}
\label{eq:Moreau}
{\mathcal M}_v[\rho](x) = \mbox{min}_{\h} \left[  \frac{(\h-x)^2}{2v}+\rho(\h)\right] 
\end{equation}
defines the so-called Moreau envelope of $\rho(\h)$, which plays an important role in convex optimisation and nonlinear analysis~\cite{Parikh:2014}. The minimum over $\h$ in the definition of the Moreau envelope~\eqref{eq:Moreau} seeks to minimise 
$\rho(\h)$ while at the same time staying close to $x$, with the relative weight of these two objectives being controlled by $v$. The Moreau envelope also appears in the context of optimal linear regression~\cite{el2013robust,bean2013optimal,donoho2013high}, where it emerges in a statistical mechanics analysis as well~\cite{advani2016statistical}. 

\para{Order parameters} The order parameters $q$ and $R$ appearing in the free energy \eqref{eq:free_energy} describe the statistics of the reconstructed couplings. At the extremum 
\eqref{eq:free_energy}, the order parameter 
\begin{equation}
\label{eq:Rdef}
R=\frac{1}{N}\sum_j\llangle  J_{1j} J_{1j}^{\star} \rrangle 
\end{equation}
 describes the (non-normalized) overlap between the reconstructed couplings and the underlying couplings. 
Similarly, the order parameter 
\begin{equation}
\label{eq:qdef}
q=\frac{1}{N}\sum_j\llangle  J_{1j} J_{1j} \rrangle
\end{equation}
 gives the overlap between a row vector of reconstructed couplings and itself. 
These order parameters turn out to be self-averaging in the thermodynamic limit: although $\frac{1}{N}\sum_j  J_{1j} J_{1j} $ fluctuates between different realisations of the couplings $\textbf{J}^{\star}$ and the samples, these fluctuations vanish with increasing system size, so for (nearly) all realisations of the disorder we have 
$\frac{1}{N}\sum_j  J_{1j} J_{1j} =\frac{1}{N}\sum_j\llangle  J_{1j} J_{1j}  \rrangle$, and similarly for the overlap $R$.

The distribution of the reconstructed couplings can also be calculated from the partition function~\eqref{eq:big_partition0}, see Supplemental Material. 
Collecting all spin pairs where the underlying coupling takes on a particular value $J^{\star}$, the corresponding reconstructed couplings turn out to follow a Gaussian distribution with mean $\frac{R J^{\star}}{q^{\star}}$ and a variance $q-R^2/q^{\star}$. For the reconstruction to have no bias, the overlap $R$ thus needs to equal the variance of the underlying couplings $q^{\star}$, for then $J$ is a random variable with mean $J^{\star}$ \footnote{In the regime of interest here, where $\alpha=M/N$ is finite, even objective functions like pseudolikelihood lead to a biased reconstruction, and become unbiased only in the limit $\alpha \to \infty$.}. 

\para{The optimal objective function}
\begin{figure}[tbh!]
\includegraphics[width = .45\textwidth]{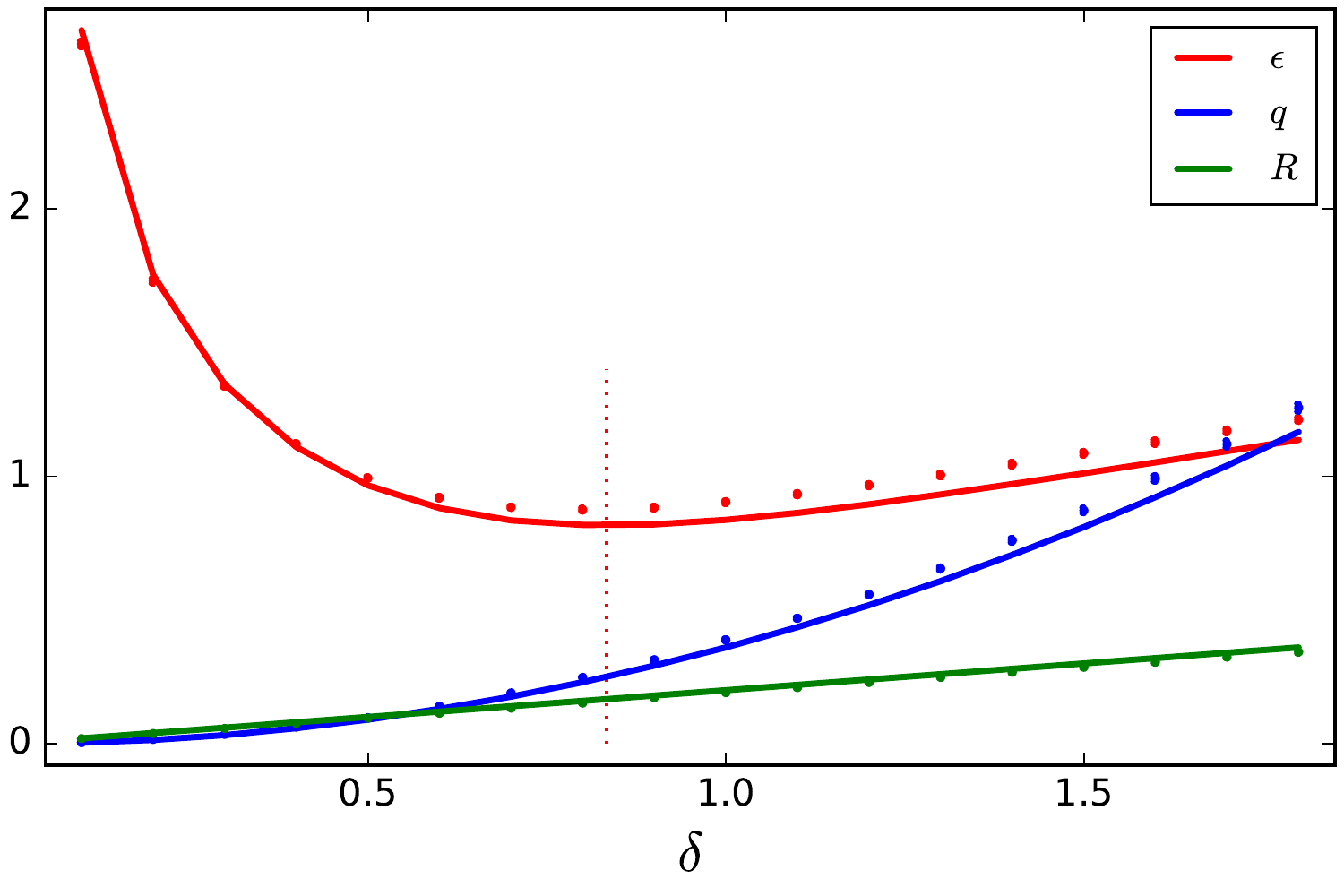}
\caption{\textbf{Reconstruction with the optimal objective function~\eqref{eq:optimal_rho}.} 
The overlaps $R$ and $q$ given by~\eqref{eq:R(delta)} and~\eqref{eq:q(delta)} and the reconstruction error $\epsilon$ defined by~\eqref{eq:relmse} are plotted against the offset $\delta$ in green, blue, and red, respectively (bottom to top). 
The numerical results were obtained by reconstructing a single system of $N=100$ spins: The underlying couplings $J^{\star}_{ij}$ were drawn independently from a Gaussian with mean zero and variance $q^{\star}=0.25$. Next, $M=500$ samples ($\alpha=M/N=5$) were generated by independent Monte Carlo runs with a breaking-in time of $100$ Monte Carlo sweeps each to ensure equilibrium had been reached. To reconstruct the couplings, the optimal objective function~\eqref{eq:optimal_rho} was minimised over separate rows of the coupling matrix using 
the \texttt{NLopt} package in Julia using Newton's method \texttt{LD\_TNEWTON}. The overlap parameters $R$ and $q$ and the reconstruction error $\epsilon$ were computed row-by-row. We plot averages over rows with the standard error as error bars (smaller than the symbol size).
The vertical line indicates the value of $\delta= (1+q^{\star}) \sqrt{\frac{q^{\star}(\alpha-1)}{\alpha q^{\star} +1}}$ which minimises the reconstruction error. 
}
\label{fig1}
\end{figure}
The two order parameters $R$ and $q$ also specify the reconstruction error. We look at the relative mean-square error
\begin{equation}
\label{eq:relmse}
\epsilon^2=
\frac{\sum_j (J_{1j}-J^{\star}_{1j})^2}{\sqrt{(\sum_j J^2_{1j}) (\sum_j  J^{*2}_{1j})}}=\frac{q-2R+q^{\star}}{\sqrt{q q^{\star}}} \ .
\end{equation}
and seek the particular objective function $\rho_{\text{opt}}(\h)$, which minimises this error. Using the calculus of variations applied to the free energy~\eqref{eq:free_energy}, 
we find 
\begin{equation}
\label{eq:optimal_rho}
\rho_{\text{opt}}(\h) = \h^2 - 2 \delta \h \ ,
\end{equation}
a square function with a non-trivial offset, whose value is $\delta= (1+q^{\star}) \sqrt{\frac{q^{\star}(\alpha-1)}{\alpha q^{\star} +1}}$, see Supplemental Material for details. Error measures different from~\eqref{eq:relmse} which also depend on the order parameters $R$ and $q$ yield the same quadratic form of the optimal objective function, but have different values of $\delta$. Finding the optimal objective function thus requires the variance $q^{\star}$ of the unknown couplings. $q^{\star}$ and hence the offset can be determined as follows: For the objective function
\eqref{eq:optimal_rho}, the free energy~\eqref{eq:free_energy} can be calculated easily, giving the overlap parameters 
\begin{align}
R&=\frac{q^{\star}\delta}{1+q^{\star}} \label{eq:R(delta)}\\
q&=\frac{(\alpha q^{\star}+1) \delta^2}{(1+q^{\star})^2(\alpha-1)} \label{eq:q(delta)}\ .
\end{align}
The overlap $q=\frac{1}{N}\sum_j J_{1j}^2$ of reconstructed couplings can be calculated easily without knowing the underlying couplings. $q^{\star}$ and thus the optimal value of the offset $\delta$ can thus be determined from a simple linear fit of $q$ against $\delta^2$. An alternative way to determine $q^{\star}$ based on spin-spin correlations in the $M$ spin configurations is 
discussed in the appendix.
In Figure \ref{fig1}, we treat the offset $\delta$ as a free parameter and show the reconstruction error as well as the overlaps $R$ and $q$ for different values of $\delta$ and compare them to numerical simulations.  

\begin{figure}[tbh!]
\includegraphics[width = .45\textwidth]{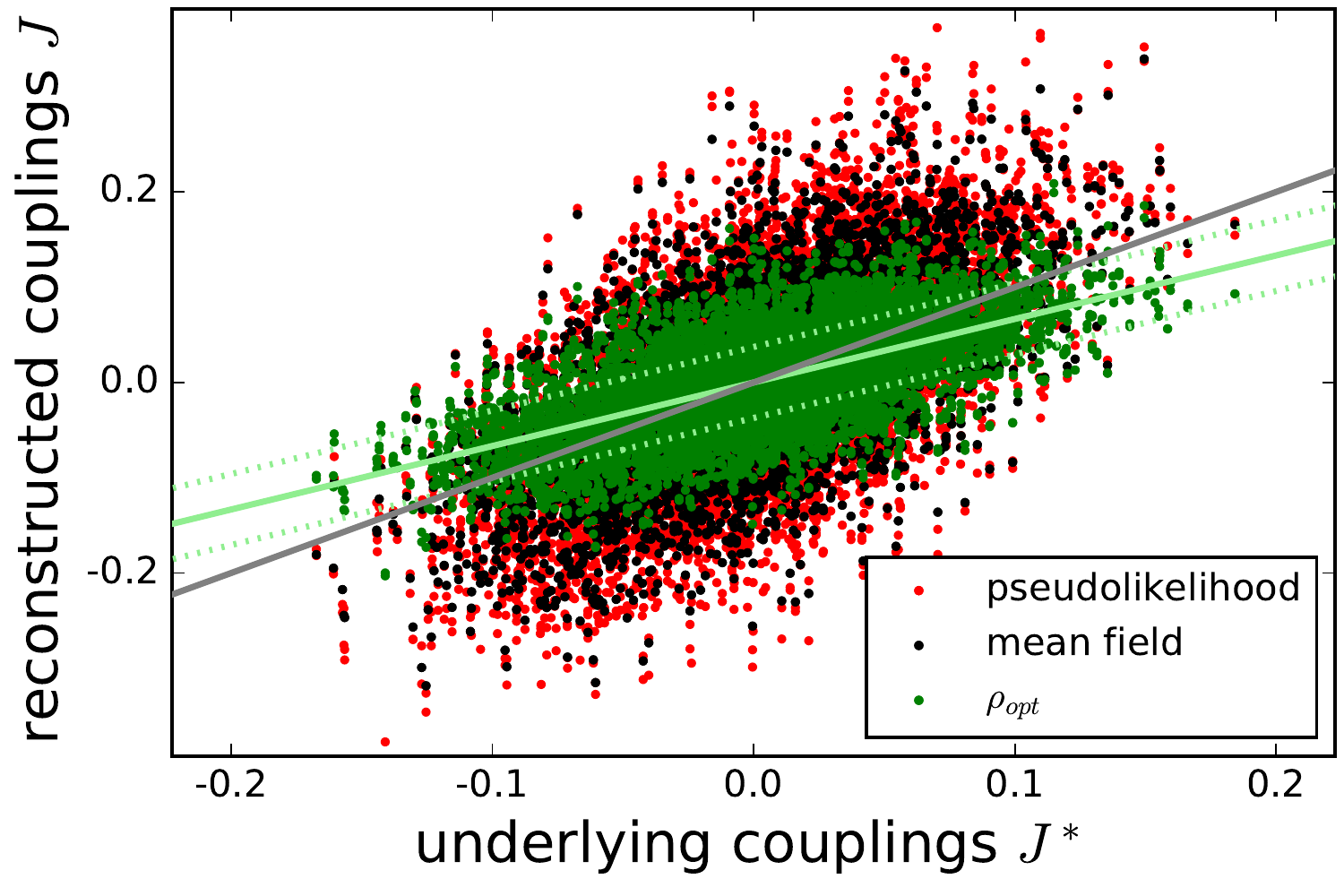}
\caption{\textbf{Reconstructed couplings versus underlying couplings.} The elements $J$ of the reconstructed coupling matrix 
are plotted against the corresponding underlying couplings $J^{\star}$. Parameters and numerical procedures are the same as in Fig.~\ref{fig1}. 
Perfect reconstruction $J=J^{\star}$ is indicated by the grey line along the diagonal. 
The results from the optimal objective function~\eqref{eq:optimal_rho} are shown in green, the reconstruction using pseudolikelihood~\eqref{eq:pl_rho}
in red. Black dots show the results of mean-field reconstruction $\textbf{J}=-{\bf \chi}^{-1}$, where ${\bf \chi}$ is the matrix of connected two-point correlations~\cite{kappen1998b}. The light green lines show the statistics of the reconstructed couplings calculated analytically; the 
solid line $J=\frac{R J^{\star}}{q^{\star}}$ gives the mean reconstructed couplings, the dotted lines are one standard deviation above and below that mean.
}
\label{fig2}
\end{figure}

\begin{figure}[tbh!]
\includegraphics[width = .45\textwidth]{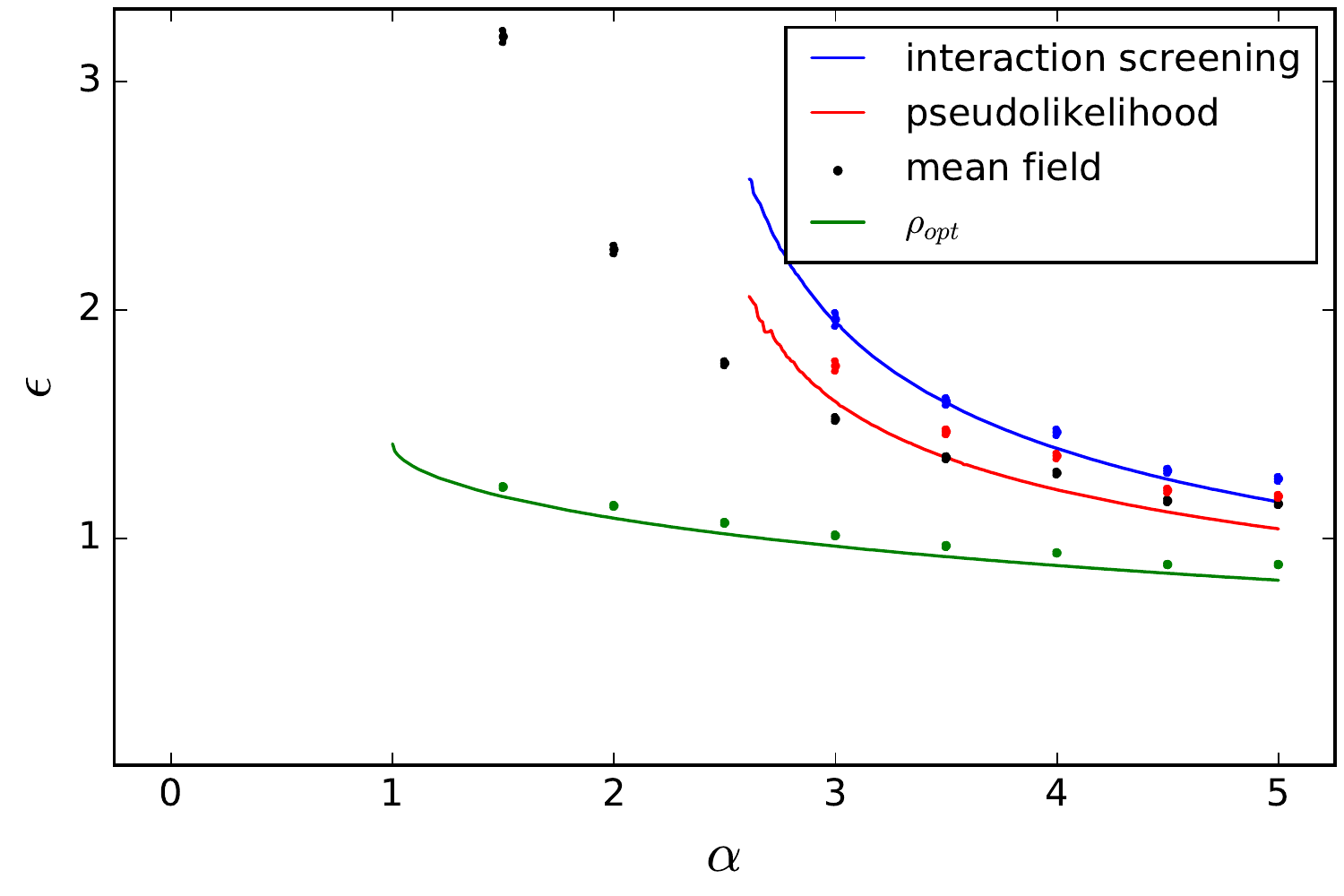}
\includegraphics[width = .46\textwidth]{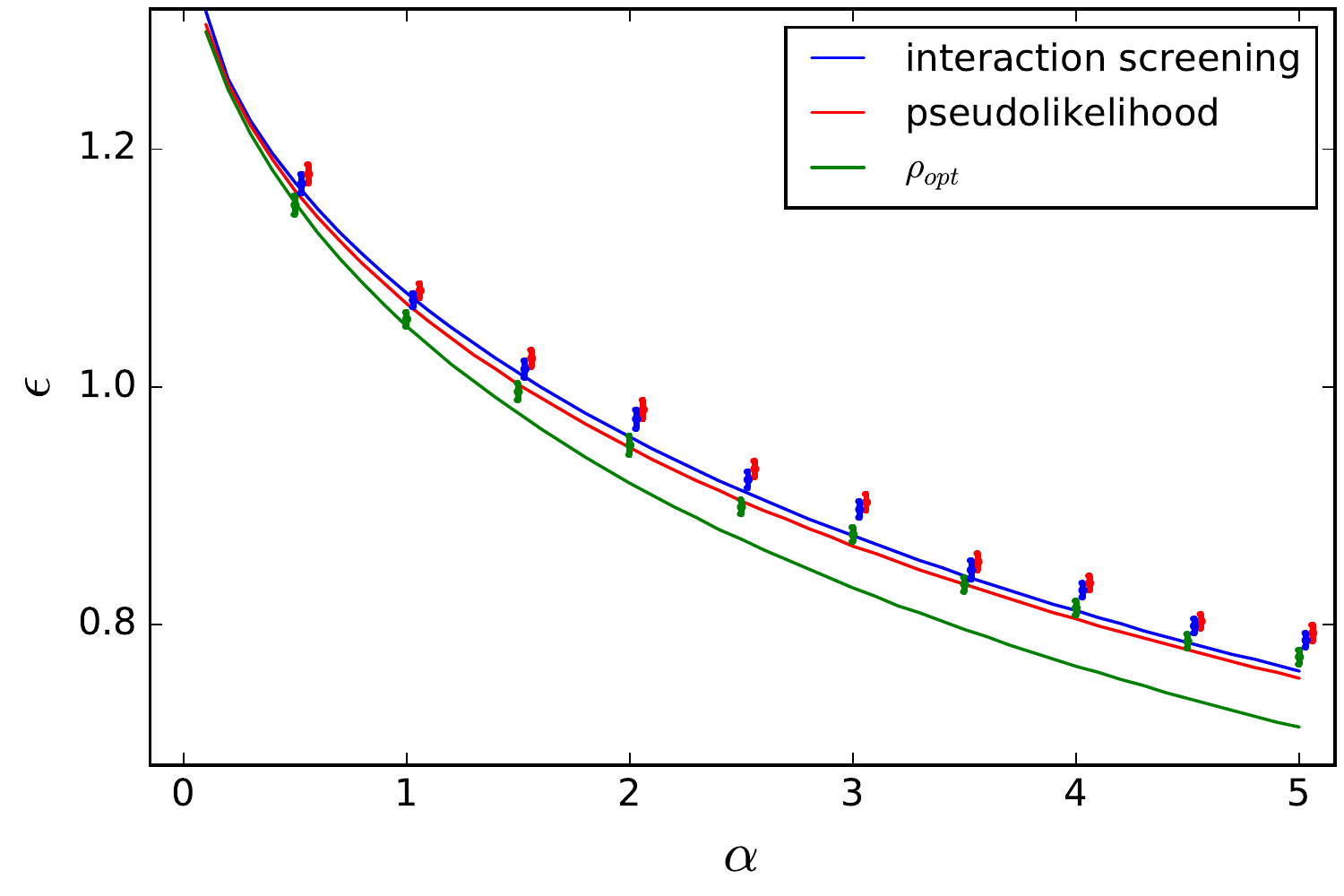}
\caption{\textbf{The reconstruction error $\epsilon$ at different numbers of samples}. 
We plot the reconstruction error $\epsilon$ defined by~\eqref{eq:relmse} against the number of samples per spin $\alpha=M/N$. The results from the optimal objective function~\eqref{eq:optimal_rho} are shown in green, pseudolikelihood~\eqref{eq:pl_rho} in red, and interaction screening~\eqref{eq:interaction_screening} in blue (bottom to top). The corresponding lines give the analytical results based on the free energy~\eqref{eq:free_energy}. Black dots indicate the results of mean-field reconstruction. Parameters and procedures are as in Figure~\ref{fig1}, except the optimisation is performed with the algorithm \texttt{LD\_MMA}, which turns out to be more stable. 
(Bottom) The same plot with with a regularizing term in the objective function~\eqref{eq:objective_reg}. The points for interaction screening and pseudolikelihood have been shifted by a small amount to the right to avoid the symbols overlapping. The value of the regularisation parameter $\gamma$ was set to match the variance of the reconstructed couplings $q$ with the variance of the underlying couplings $q^{\star}$ for each row of the coupling matrix, see Supplemental Material.  
}
\label{fig3}
\end{figure}

Figure~\ref{fig2} compares reconstructed and underlying couplings for different methods; the optimal objective function~\eqref{eq:optimal_rho}, pseudolikelihood~\eqref{eq:pl_rho}, and mean-field reconstruction~\cite{kappen1998b}, showing that the optimal objective function~\eqref{eq:optimal_rho} outperforms both of these methods. Figure~\ref{fig3} (top) compares the reconstruction error $\epsilon$ for these three methods, as well as interaction screening~\eqref{eq:interaction_screening}, at different values of $\alpha=M/N$.  The optimal objective function performs best, with a particularly wide margin at low values of $\alpha$. The reconstruction error increases for all four methods as $\alpha$ decreases, most rapidly for pseudolikelihood, interaction screening, and mean-field reconstruction. For mean-field reconstruction, the rapid increase of the reconstruction error with decreasing $\alpha$ is connected to the matrix of two-point spin correlations becoming singular at $\alpha=1$.  For mean-field reconstruction, but also for the reconstruction based on pseudolikelihood and interaction screening, we find that the self-overlap parameter $q$ diverges as $\alpha$ approaches one from above. For $\alpha$ below three, the convex optimisation algorithms fail for pseudolikelihood and interaction screening and also the numerical extremization of the free energy~\eqref{eq:free_energy} fails. 

\para{Reconstruction with a regularizer}
For pseudolikelihood and interaction screening this divergence can be avoided by adding a regularizing term to the objective function. 
Regularizing terms are often used to control the sparsity of the coupling matrix. Here we use a quadratic regularizer leading to the objective function
\begin{equation}
\label{eq:objective_reg}
\sum_{\mu} \rho(\h_1^{\mu}) + \frac{\gamma}{2} \sum_j J_{1j}^2 \ .
\end{equation}
This regularisation term penalizes large values of the couplings, so 
the regularisation parameter $\gamma$ can be used to control the self-overlap $q$ of the reconstructed couplings. 
The objective function~\eqref{eq:objective_reg} yields the same free energy as~\eqref{eq:free_energy} above, except for an additional term $-\gamma q/2$, see Supplemental Material. 
The value of $\gamma$ can be determined from data in the same way the value of the offset $\delta$ was determined; by re-calculating the overlap parameters $q$ and $R$ from the free energy, matching the dependence of $q$ on $\gamma$ with numerical results to determine $q^{\star}$, and then solving $q(\gamma)=q^{\star}$ for the regularisation parameter. Similarly, we calculate the optimal $\rho(\h)$ in the presence of a regularizer and find again the quadratic function
\eqref{eq:optimal_rho}, but a different optimal value of $\delta$ (see Supplemental Material). Figure~\ref{fig3} (bottom) compares the reconstruction error $\epsilon$ for the optimal $\rho(\h)$, pseudolikelihood, and interaction screening in the presence of the regularizing term. It shows that adding the regularizing term allows the extension of all three methods to values of $\alpha$ below one. The reconstruction errors of pseudolikelihood and interaction screening are very close to each other, and while the reconstruction error of the optimal $\rho(\h)$ is always smaller than that of the other methods, the difference is only in the range of  $2-3\%$ in numerical simulations with $N=100$. Thus the performance of both pseudolikelihood and interaction screening with a regularizer is close to optimal in the regime probed here. 

\para{Conclusion}
In the inverse Ising problem, one infers the parameters of an Ising model on the basis of spin configurations drawn from the equilibrium distribution. This is the reverse direction compared to the standard statistical mechanics problem, the so-called forward problem, where observables like correlations and magnetisations are calculated given the model parameters. In this paper, we have used an inverse statistical mechanics to match this reversal of direction: the partition function~\eqref{eq:big_partition0} has the couplings between spins as degrees of freedom, whereas the spin configurations are drawn once from the Boltzmann distribution and then remain fixed (quenched disorder). Such a reversal of direction has been made before in a different context, namely the statistical mechanics of neural network~\cite{watkin1993a,engel2001a}. We have applied this approach to a simple scenario characterized by fully-connected coupling matrices, for which we analytically calculated the optimal objective function.

Several open questions remain, some of them of a technical nature like the validity of the replica-symmetric ansatz used to calculate the free energy~\eqref{eq:free_energy}.  The objective function~\eqref{eq:objective_reg} is a convex function of the couplings. As a result, any local minimum of the objective function is also a global minimum, so we do not expect a spontaneous breaking of replica symmetry (describing a situation with multiple minima).  However, we expect the small-coupling resummation used to derive the free energy to fail for large coupling strengths, see appendix. Bachschmid-Romano and Opper~\cite{bachschmid2017statistical} have recently analysed the inverse Ising problem using the cavity approach~\cite{mezardparisivirasoro}. Their approach allows to circumvent the high-temperature expansion used here. While the results of Bachschmid-Romano and Opper agree with our results at high temperatures (low $q^{\star}$), they differ at low temperatures and show very good agreement with numerical results. Specifically, their free energy essentially agrees with our result~\eqref{eq:free_energy}, but the interpretation of the order parameters differs. A consequence is the discrepancy between analytical and numerical results in figures~\ref{fig1} and~\ref{fig3}, which grows with increasing $q^{\star}$. In the appendix we re-derive their result (without a regularizing term) using our expansion, leaving the combination with a regularizing term for future work.

Another point is the reconstruction error~\eqref{eq:relmse}, which is based on the Euclidian distance between the underlying and the reconstructed couplings. (This is different from the regularizing term in~\eqref{eq:objective_reg}.) For the reconstruction error, several alternative choices will be interesting. One of them is using the $\ell_0$-norm, which count links between spins with non-zero couplings differing between the original and the reconstructed systems~\cite{vuffray2016interaction,lokhov2016optimal}. Alternatively, one may focus on couplings with large absolute values: In practice, frequently the $k$ spins pairs with the largest couplings are retained for comparison to the underlying couplings, or to the results of other reconstruction methods. The focus on couplings with large values could be implemented by an $\ell_p$-reconstruction error; $\epsilon^p= \frac{\sum_j (J_{1j}-J^{\star}_{1j})^p}{\sqrt{(\sum_j J^p_{1j}) (\sum_j J^{*p}_{1j})}}$ with a large even value of $p$.

Finally, an important scenario to consider is sparse coupling matrices. In practice, coupling matrices are often sparse, and reconstruction requires a regularizing term such as ~\eqref{eq:objective_reg}. Spin glasses with sparse couplings (diluted spin glasses) are characterized by a non-trivial distribution of the effective local fields, which the optimal objective function in combination with a regularizing term could exploit.

{\bf Acknowledgments:} Many thanks to Guy Bunin, David Gross, Ulrich Michel, and Chau Nguyen for discussions.

\bibliography{inv-ising}

\newpage
\onecolumngrid
\appendix
\newpage
\section*{Statistical mechanics of the inverse Ising problem and the optimal objective function: Supplemental Material}

\subsection*{1. Computing the partition function}
To average $\ln Z (\bf{s}^1,\bf{s}^2,\ldots,\bf{s}^M)$ in \eqref{eq:big_partition0} over the disorder, we use the replica trick in two separate instances. First,  to represent the logarithm of the partition function 
in~\eqref{eq:partitionJopt}, we use $\ln Z = \lim_{n \to 0} \partial_n Z^n$. The inverse of the  partition function in 
Boltzmann distribution~\eqref{eq:boltzmann} is represented with a second set of replicas based on 
\begin{align}
\frac{\sum_{\textbf{s}} e^{-\beta {\mathcal H(\textbf{s})}} f(\textbf{s}) }{\sum_{\textbf{s}} e^{-\beta {\mathcal H}(\textbf{s})}} = \lim_{m \to 0} \prod_{\alpha=1}^{m}(\sum_{\textbf{s}^{\alpha}})
e^{-\beta \sum_{\alpha}{\mathcal H}(\textbf{s}^{\alpha})}f(\textbf{s}^1)\ .
\end{align}
Taking the underlying couplings $J^{\star}_{ij}$ to be taken independently from a Gaussian distribution with zero mean 
and variance $q^{\star}$ the average of $Z^n(\bf{s}^1,\ldots,\bf{s}^M)$ is 
\begin{align}
\label{eq:big_partition1}
\llangle &Z^n({\bf{s}}^1,\ldots,{\bf{s}}^M) \rrangle= \prod_{i<j} \int \frac{dJ_{ij}^{\star}}{\sqrt{2 \pi q^{\star}}} \exp\left\{ - \frac{1}{2q^{\star}} \sum_{i<j} J_{ij}^{\star 2} \right\}\nonumber\\
&\prod_{\mu,\alpha}\left( \frac{1}{2^N} \sum_{\textbf{s}^{\mu \alpha}}\right) \exp\left\{ \sum_{i<j,\mu,\alpha} \frac{ J_{ij}^{\star}}{\sqrt{N}} s_i^{\mu \alpha} s_j^{\mu \alpha} \right\}\\
&\prod_a \left(\int d\textbf{J}^a\right) \prod_{\mu,a} \left(\int \frac{d\h^{\mu\,a}d\hat{\h}^{\mu\,a}}{2\pi}\right) \exp\left\{ -i \sum_{\mu\, a}\h^{\mu\,a}\hat{\h}^{\mu\,a} \right\} 
\nonumber\\
&  \exp\left\{ \frac{i}{\sqrt{N}} \sum_{\mu\, a}\hat{\h}^{\mu\,a} s_1^{\mu 1}  \sum_{j \neq 1} J_j^a s_j^{\mu\,1}  -\beta \sum_{\mu\,a} \rho(\h^{\mu\,a}) \right\} \nonumber\ ,
\end{align}
where the replica indices $a$ and $\alpha$ run from $1$ to $n$ and $1$ to $m$ respectively, and the limits $m\to 0$ and $n\to 0$ will be taken at the end of the calculation. 
The vector $\bf{J}$ with elements $J_j=J_{1j}$ is the first row of the matrix of inferred couplings. We have inserted a factor of $1/2^N$ into the partition functions for convenience, 
in the limit of $m \to 0$ they will cancel each other out. 
We have used a set of delta-functions to define the argument of the objective function, $\h^{\mu} = s_1^{\mu} \sum_{j\neq i} J_j s_j^{\mu}$. Partition functions of this form have been investigate extensively in the context of statistical learning~\cite{watkin1993a,engel2001a}. The key difference here is that the samples $\bf{s}^1,\ldots,\bf{s}^M $ are not generated from some `teacher perceptron', but are taken from the Boltzmann distribution~\eqref{eq:boltzmann}. 

\begin{figure}[tbh!]
\includegraphics[width = .4\textwidth]{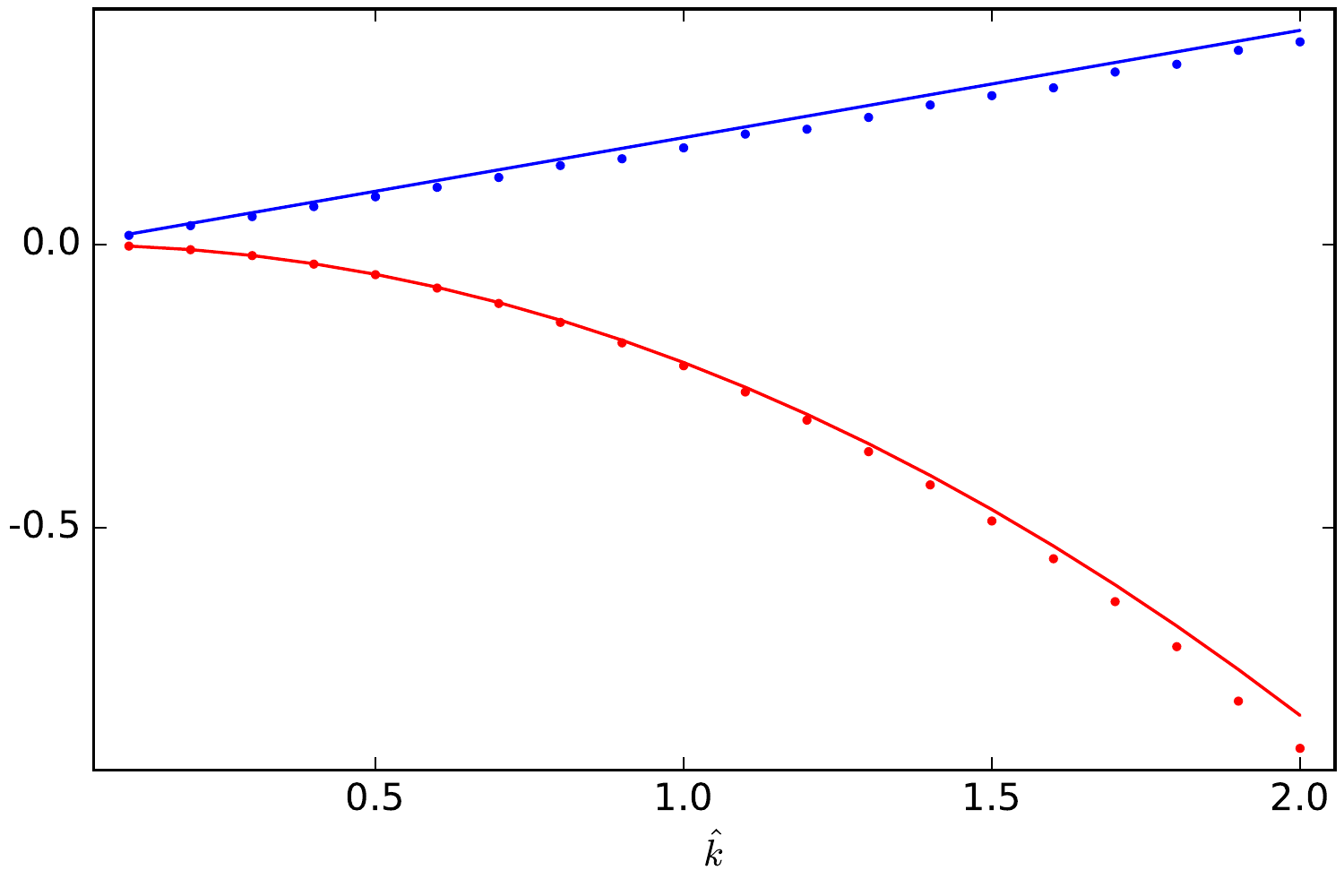}
\caption{\textbf{The average over samples.} 
We compare the logarithm of the right and left hand side of the product of equations  \eqref{eq:patternaverage_alpha1} - \eqref{eq:patternaverage_alphag1} as a function of the parameter $\hat{\h}$ for $n=1$ and $m=0$. $J^{\star}_{ij}$ with $i<j$ are i.i.d. Gaussian entries of zero mean and variance $q^{\star}=0.25$ and $J_j=J^{\star}_{1j}+x_j$, where the $x_j$ are also i.i.d. Gaussian random variables of zero mean and variance $q^{\star}$.  $J_1$ is set to zero as it corresponds to a self-interaction. The numerical averages on the left hand sides are indicated by points, the analytical results on the right hand sides are described by lines. As the product of  \eqref{eq:patternaverage_alpha1} and \eqref{eq:patternaverage_alphag1}  is complex, real parts are shown in red (bottom), imaginary parts in blue (top). 
\label{fig_average} 
}
\end{figure}

The partition function~\eqref{eq:big_partition1} can be evaluated by standard techniques~\cite{watkin1993a,engel2001a}, except for the first step, the sum over the samples $s_i^{\mu\,\alpha}$. Picking out the terms involving the samples, the average factorises over the sample index $\mu$ (which we drop in the following for convenience) leaving
\begin{equation}
\prod_{\alpha}\left( \frac{1}{2^N} \sum_{\textbf{s}^{\alpha}}\right) \exp\left\{ \sum_{i<j,\alpha} \frac{ J_{ij}^{\star}}{\sqrt{N}} s_i^{ \alpha} s_j^{\alpha} 
+\frac{i}{\sqrt{N}} s_1^1 \sum_{a}\hat{\h}^{a}\sum_j J_j^a s_j^{1}\right\} \ .
\end{equation}
The contribution from $\alpha=1$ is 
\begin{equation}
\frac{1}{2^N} \sum_{\textbf{s}^1}\exp\left\{ \frac{1}{\sqrt{N}}  \sum_{i<j}   s_1^1( J_{ij}^{\star} +  i \sum_a \hat{\h}^{a} J_j^a  \delta_{i1}) s_j^{1} \right\} 
= \prod_i^N \left(\frac{1}{2} \sum_{s_i^1} \right) \exp^{\frac{1}{\sqrt{N}} \sum_{i<j} s_i^1 G_{ij} s_j^1} \ ,
\end{equation}
where we have introduced the shorthand $G_{ij}= J_{ij}^{\star} +  i \sum_a \hat{\h}^{a} J_j^a  \delta_{i1}$. 
Expanding the exponent in a Taylor series (small couplings, \textit{i.e.}, small $q^{\star}$, or high temperatures) gives 
\begin{equation}
\label{eq:hTexpansion}
\prod_i^N \left( \frac{1}{2}\sum_{s_i^1} \right) \left[ 1+ \frac{1}{\sqrt{N}} \sum_{i<j} s_i G_{ij} s_j +\frac{1}{2!}(\frac{1}{\sqrt{N}})^2  \sum_{i<j} s_i G_{ij} s_j  \sum_{k<l} s_k G_{kl} s_l + \ldots  \right] \ .
\end{equation}
The first-order term in this expansion sums to zero, to yield a non-zero result would require $i=j$, which does not appear in the sum.  
For the second order expression, terms with $i=k$, $j=l$ sum to $\frac{1}{2!N} \sum_{i<j} G_{ij}^2$, other terms either sum to zero or are smaller by a factor of $N^{-1/2}$. For higher-order terms, the dominant contributions come from terms where spin pairs are contracted in the same manner as in the second-order term.  
The dominant (2n)-th order term is $\frac{1}{(2n)!N^n} [(2n-1)(2n-3)\ldots 3\times 1](\sum_{i<j} G_{ij}^2)^n=\frac{1}{n!2^nN^n} (\sum_{i<j} G_{ij}^2)^n $, resumming the series to infinite order
 gives 
\begin{equation}
\label{eq:patternaverage_alpha1}
 \prod_i^N \left( \frac{1}{2} \sum_{s_i^1} \right) \exp^{\frac{1}{\sqrt{N}} \sum_{i<j} s_i^1 G_{ij} s_j^1} = \exp\left\{ \frac{1}{2N}  \sum_{i<j} G_{ij}^2\right\} \ .
\end{equation}
Expanding the shorthand in this result we have 
\begin{equation}
\label{eq:s_average_exponent}
\frac{1}{2N}\sum_{i,j} G_{ij}^2= \frac{1}{2N}\sum_{i<j} J_{ij}^{\star 2} + \frac{i}{N} \sum_a \hat{\h}^{a} \sum_j J_{1j}^{\star} J_j^a - \frac{1}{2N} \sum_{a,b} \hat{\h}^a \hat{\h}^b \sum_j J^a_j J^b_j 
\end{equation}
An analogous calculation can be made for the terms with $\alpha>1$ which gives
\begin{equation}
\label{eq:patternaverage_alphag1}
\prod_{i=1,\alpha=2}^{N,m}\left(\frac{1}{N} \sum_{s_i^{\alpha}} \right) \exp^{\frac{1}{\sqrt{N}} \sum_{i<j\alpha=2} s_i^{\alpha} J^{\star}_{ij} s_j^{\alpha} }=\exp\left\{\frac{m-1}{2N} \sum_{i<j} J_{ij}^{\star 2}\right\} \ .
\end{equation}
In the limit $m \to 0$ this term cancels with the first term of~\eqref{eq:s_average_exponent}. Note that the first term in  
~\eqref{eq:s_average_exponent} scales differently with $N$ from the remaining terms. In order to probe this result numerically at finite $N$, we compute the averages on the left hand sides \eqref{eq:patternaverage_alpha1} and \eqref{eq:patternaverage_alphag1} for a single matrix  numerically and compare the logarithm of their product (so the first term in~\eqref{eq:s_average_exponent} cancels with~\eqref{eq:patternaverage_alphag1}) with the analytical result $+ \frac{i}{N} \hat{\h} \sum_j J_{1j}^{\star} J_j- \frac{1}{2N} \hat{\h}^2  \sum_j J_j J_j$. For $N=20$, figure \ref{fig_average} compares the numerical average over $2^N$ samples with the analytical result. 

The remaining terms in~\eqref{eq:s_average_exponent} can be simplified by introducing the order parameters $q_{ab}=\frac{1}{N} \sum_i J_i^a J_i^b$ and $R_a=\frac{1}{N} \sum_i J_i^a J_{1i}^{\star}$ via integrals over delta-functions. 
Thus we obtain in the limit $m \to 0$
\begin{align}
\label{eq:big_partition2}
\llangle &Z^n(\textbf{s}^1,\ldots,\textbf{s}^M) \rrangle= \prod_{i<j} \int \frac{dJ_{ij}^{\star}}{\sqrt{2 \pi q^{\star}}} \exp\left\{ - \frac{1}{2q^{\star}} \sum_{i<j} J_{ij}^{\star 2} \right\}\\
&\prod_{a \leq b} \int \frac{d q_{ab}d \hat{q}_{ab}}{2 \pi/N}
\prod_{a}  \int \frac{d R_a d \hat{R}_{a}}{2 \pi/N} \nonumber
\exp\{-iN\sum_{a \leq b} q_{ab}\hat{q}_{ab}  -i \sum_a R_a \hat{R}_a  \} \nonumber\\
&\prod_a \int d\textbf{J}^a \exp\{+i \sum_{a\leq b} \hat{q}_{ab} \sum_i J_i^a J_i^b+i\sum_a \hat{R}_a \sum_i J_i^a J^{\star}_{1i}\}  \nonumber\\
&\prod_{a,\mu} \int \frac{d\h^{\mu\,a}d\hat{\h}^{\mu\,a}}{2\pi} \exp\{ -i\sum_{\mu a} \h^{\mu\,a}\hat{\h}^{\mu\,a} +i \sum_{\mu a}\hat{\h}^{\mu\,a}R_a
-\frac{1}{2} \sum_{a,b,\mu} \hat{\h}^{\mu\,a}\hat{\h}^{\mu\,b} q_{ab} -\beta \sum_{a,\mu} \rho(\h^{\mu\, a}) \}\nonumber
\end{align}
where the $\delta$-functions themselves were represented by integrals over the so-called 
conjugate order parameters $\hat{q}_{ab}$ and $\hat{R}_a$. In the next step, we exploit that the integrals over couplings factorise over $i=1,\ldots,N$ and those over variables $\h^{\mu\,a}$ and $\hat{\h}^{\mu\,a}$ factorise over  $\mu=1,\ldots,M=\alpha N$, giving
\begin{equation}
\label{eq:big_partition3}
\llangle Z^n(\textbf{s}^1,\ldots,\textbf{s}^M) \rrangle = 
\prod_{a \leq b} \int \frac{d q_{ab}d \hat{q}_{ab}}{2 \pi/N}
\prod_{a}  \int \frac{d R_a d \hat{R}_{a}}{2 \pi/N} \nonumber
\exp\{-iN\sum_{a \leq b} q_{ab}\hat{q}_{ab}  -i N\sum_a R_a \hat{R}_a  +Ng_S(\{\hat{q}_{ab},\hat{R}_a\}) +\alpha N g_E(\{q_{ab},R_a\}) \} 
\end{equation}
with 
\begin{align}
\label{eq:gdef}
e^{g_S(\{\hat{q}_{ab},\hat{R}_a\})} &= \int D_{q^{\star}}J^{\star} \prod_a \int dJ^a \exp \{ i \sum_{a \leq b} \hat{q}_{ab} J^a J^b + i \sum_a \hat{R}_a J^a J^{\star}   \} \\
e^{g_E(\{q_{ab},R_a\})} &= \prod_a \int \frac{d\h^a d \hat{\h}^a}{2 \pi } \exp\{  -i\sum_a \h^a \hat{\h}^a 
+ i \sum_a \hat{\h} ^aR_a -\frac{1}{2} \sum_{a,b} \hat{\h}^a  \hat{\h}^b q_{ab} -\beta \sum_a \rho(\h^a)  \}
\end{align}
where $D_q x = \frac{dx}{\sqrt{2\pi q}} e^{-x^2/(2q)}$ denotes a Gaussian measure with mean zero and variance $q$. At this point we take a replica symmetric ansatz defined by 
\begin{align}
q_{aa}=q_1 \ \ \ \ & i \hat{q}_{aa} = -\frac{1}{2} \hat{q}_1 & \forall a \\
q_{ab}=q_0 \ \ \ \  & i \hat{q}_{ab} =  \hat{q}_0 & \forall a < b \nonumber\\ 
R_a=R      \ \ \ \    & i\hat{R}_a=\hat{R} & \forall a \nonumber
\end{align} 
which allows the evaluation of \eqref{eq:big_partition3} and the taking of the limit $n\to 0$ yielding 
\begin{align}
\label{eq:big_partition4}
-\beta f &\equiv  \frac{1}{N}\llangle \ln Z(\textbf{s}^1,\ldots,\textbf{s}^M) \rrangle= \frac{1}{N}\lim_{n \to 0} \partial_n \llangle Z^n(\textbf{s}^1,\ldots,\textbf{s}^M) \rrangle = \mbox{extr}_{q_1,\hat{q}_1,q_0,\hat{q}_0,R,\hat{R}} 
\left[ \frac{1}{2} q_1 \hat{q}_1 + \frac{1}{2} q_0 \hat{q}_0 -R \hat{R} \right. \nonumber\\
&\left.+ \frac{1}{2} \ln(2\pi) -  \frac{1}{2} \ln(\hat{q}_1+\hat{q}_0) 
+\frac{1}{2} \frac{\hat{q}_0+q^{\star} \hat{R}^2}{\hat{q}_1+\hat{q}_0} 
+\alpha \int Dt \ln \left[ \int \frac{d\h}{\sqrt{2\pi(q_1-q_0)}} \exp\{- \frac{(\h-R-\sqrt{q_0}t)^2}{2(q_1-q_0)}  -\beta \rho(\h)  \}
\right]
\right] \ .
\end{align}
The extremum over the conjugate order parameters $\hat{q}_1,\hat{q}_0,\hat{R}$ can be evaluated easily yielding 
\begin{align}
\label{eq:conjugate_ops}
\hat{q}_1+\hat{q}_0 &= \frac{1}{q_1-q_0} \\
\hat{R}&=\frac{R}{q^{\star}(q_1-q_0)}\\
\hat{q}_1&=\frac{q_1-2 q_0+R^2/q^{\star}}{(q_1-q_0)^2}
\end{align}
which gives 
\begin{align}
\label{eq:big_partition5}
-\beta f = \mbox{extr}_{q_1,q_0,\hat{q}_0,R} \left[ 
\frac{1}{2} \frac{q_1- R^2/q^{\star}}{q_1-q_0} + \frac{1}{2} \ln(q_1-q_0)  + \frac{1}{2} \ln(2\pi) + \alpha \int Dt \ln \left[ \int \frac{d\h}{\sqrt{2\pi(q_1-q_0)}} \exp\{- \frac{(\h-R-\sqrt{q_0}t)^2}{2(q_1-q_0)}  -\beta \rho(\h)  \}\right]
\right] \ .
\end{align}
We are particularly interested in the low-temperature limit $\beta \to \infty$; according to \eqref{eq:partitionJopt}, this limit projects out the couplings 
minimizing the objective function. In the low-temperature limit we find that the order parameters at the extremum~\eqref{eq:big_partition5} scale as $q_1-q_0 \to v/\beta$, where $v$ and $q_0$ are of order one. With this scaling, the integral over $\h$ in~\eqref{eq:big_partition5} can be evaluated by saddle-point integration. To leading order in $\beta$ we have 
  \begin{align}
\label{eq:h_saddle}
\ln \left[ \int \frac{d\h}{\sqrt{2\pi(q_1-q_0)}} \exp\{- \frac{(\h-R-\sqrt{q_0}t)^2}{2(q_1-q_0)}  -\beta \rho(\h)  \}\right] =
 -\beta \mbox{min}_{\h} \left[ \frac{(\h-R-\sqrt{q_0}t)^2}{2 v} + \rho(\h)\right] \ .
\end{align}
The minimum over $\h$ admits a simple interpretation: As a function of $x=R+\sqrt{q_0}t$, this is a minimum of $\rho(\h)$ that is `close' to $x$, where the trade-off between closeness and smallness is controlled by $v$. This relationship plays a central role in convex optimisation~\cite{Parikh:2014}, where it is known as the Moreau envelope ${\mathcal M}_v$ of a function $f$
\begin{equation}
\label{eq:Moreau_appendix}
{\mathcal M}_v[f](x) = \mbox{min}_y \left[  \frac{(y-x)^2}{2v}+f(y)\right] \ .
\end{equation}
With this we obtain
\begin{equation}
- f = \mbox{extr}_{q,v,R} \left[\frac{q-R^2/q^{\star}}{2v}  - \alpha \int Dt  {\mathcal M}_v[\rho](R+\sqrt{q}t) \right] \ .
\end{equation}
Setting the derivatives of this expression with respect to $q,v$ and $R$ to zero gives the three saddle-point equations
\begin{align}
\label{eq:spe}
\frac{1}{v} - \frac{\alpha}{\sqrt{q}} \int Dt\, t \frac{d\rho}{d\h}\vert_{\h=\h(R+\sqrt{q}t,v)} &=0 \\
-\frac{R}{q^{\star}v}- \alpha \int Dt\, \frac{d\rho}{d\h}\vert_{\h=\h(R+\sqrt{q}t,v)} &=0 \nonumber\\
-\frac{q-R^2/q^{\star}}{v^2} + \alpha \int Dt\, (\frac{d\rho}{d\h}\vert_{\h=\h(R+\sqrt{q}t,v)})^2 &=0 \nonumber\ ,
\end{align}
where $\h(R+\sqrt{q}t,v)$ is value of $\h$ attaining the minimum in \eqref{eq:h_saddle}. To derive these equations we used 
$\partial_x{\mathcal M}_v[\rho](x)=\frac{d\rho}{d\h}\vert_{\h=\h(x)}$ and $\partial_v{\mathcal M}_v[\rho](x)=-\frac{1}{2}(\frac{d\rho}{d\h}\vert_{\h=\h(x)})^2$. 

\subsection*{2. Finding the optimal objective function}

We are interested in the particular objective function that when used to reconstruct couplings according to \eqref{eq:partitionJopt}, yields reconstructed couplings that are 
closest to the underlying couplings. We use the relative mean-square error~\eqref{eq:relmse} to quantify the performance of a particular objective function. 
We thus seek the particular function $\rho(\h)$ which maximises $\frac{q-2R+q^{\star}}{\sqrt{q q^{\star}}}$, subject to the constraints 
\eqref{eq:spe} specified by the saddle point equations. A similar calculation appears in the context of optimal regression~\cite{advani2016statistical}. We use Lagrange multipliers and maximise
\begin{equation}
\label{eq:lagrange1}
L=\frac{q-2R+q^{\star}}{\sqrt{q q^{\star}}} + \gamma_1 \left[ \frac{\sqrt{q}}{v} - \alpha\int Dt\, t \rho'\right] + \gamma_2 \left[ \frac{R}{q^{\star}v}+ \alpha \int Dt\,\rho' \right] +\gamma_3 \left[\frac{q-R^2/q^{\star}}{v^2} - \alpha \int Dt\, (\rho')^2 \right] \ ,
\end{equation}
where we use the shorthand $\rho'=\frac{d\rho}{d\h}\vert_{\h=\h(x)}$. With $\rho'=\partial_x {\mathcal M}_v[\rho](x)$ we have $\int Dt \rho' = 
\int D_{R,q}x {\mathcal M}_v'(x)$, $\int Dt\, t\rho'= \sqrt{q} \int D_{R,q}x {\mathcal M}_v''(x)$ and $\int Dt (\rho')^2 =\int D_{R,q}x ( {\mathcal M}_v'(x))^2$ we can write \eqref{eq:lagrange1} as 
\begin{equation}
\label{eq:lagrange2}
L=\frac{q-2R+q^{\star}}{\sqrt{q q^{\star}}}+ \gamma_1 \frac{\sqrt{q}}{v} + \gamma_2  \frac{R}{q^{\star}v}+\gamma_3 \frac{q-R^2/q^{\star}}{v^2} + \alpha \int_{-\infty}^{\infty} D_{R,q}x {\mathcal L}(x)
\end{equation}
with
\begin{equation}
\label{eq:lagrangiandensity1}
 {\mathcal L}(x) = -\gamma_1 \sqrt{q} \frac{d^2{\mathcal M}_v}{dx^2} + \gamma_2 \frac{d{\mathcal M}_v}{dx} - \gamma_3 (\frac{d{\mathcal M}_v}{dx})^2 \ 
\end{equation}
and $D_{R,q}x$ a shorthand for a Gaussian integral measure with mean $R$ and variance $q$. To find the optimal $\rho(\h)$ we take the functional derivative of $L$ with respect to $\frac{d{\mathcal M}_v}{dx}$, solve the resulting Euler-Lagrange equation, and determine the corresponding $\rho(\h)$ by inverting the Moreau envelope
\eqref{eq:Moreau_appendix}. The Euler-Lagrange equation 
\begin{equation}
\frac{\partial}{\partial {\mathcal M}_v'} \left( G_{R,q}(x) {\mathcal L}(x) \right) - \frac{d}{dx} \frac{\partial}{\partial {\mathcal M}_v''} \left( G_{R,q}(x) {\mathcal L}(x) \right) = 0
\end{equation}
gives 
\begin{equation}
\label{eq:Mprime}
\frac{d{\mathcal M}_v}{dx}= \frac{1}{2 \gamma_3} \left( \gamma_2 + \gamma_1\sqrt{q} \frac{d}{dx} \ln G_{R,q}(x) \right)
\ ,
\end{equation}
 where $G_{R,q}(x)$ is a Gaussian with mean $R$ and variance $q$. Inserting this result into \eqref{eq:lagrangiandensity1}, gives 
\begin{equation}
\label{eq:lagrangiandensity2}
 {\mathcal L}(x) = -\frac{\gamma_1^2 q G_{R,q}'' }{2 \gamma_3 } + \frac{\gamma_1^2 q G_{R,q}'^2}{4 \gamma_3 G_{R,q}^2} + \frac{\gamma_2^2}{4 \gamma_3} \ .
\end{equation}
The first term integrates to zero, the second involves $\int_{-\infty}^{\infty} dx\, \frac{G_{R,q}'^2(x)}{G_{R,q}(x)}=\frac{1}{q}$ which gives the Lagrangian \eqref{eq:lagrange2} as
\begin{equation}
\label{eq:lagrange3}
L=\frac{q-2R+q^{\star}}{\sqrt{q q^{\star}}}+ \gamma_1 \frac{\sqrt{q}}{v} + \gamma_2  \frac{R}{q^{\star}v}+\gamma_3 \frac{q-R^2/q^{\star}}{v^2} + \alpha \frac{\gamma_1^2+\gamma_2^2}{4 \gamma_3} \ .
\end{equation}
Extremization with respect to the Lagrange parameters gives
\begin{align}
\frac{\sqrt{q}}{v}+ \frac{\alpha}{2}\frac{\gamma_1}{\gamma_3}&=0 \\
\frac{R}{vq^{\star}}+ \frac{\alpha}{2}\frac{\gamma_2}{\gamma_3}&=0 \\
\frac{q-R^2/q^{\star}}{v^2}- \frac{\alpha}{4}\frac{\gamma_1^2+\gamma_2^2}{\gamma_3^2}&=0 \ .
\end{align}
Only the first two of these equations are required to evaluate \eqref{eq:Mprime}, the third establishes a relationship between the overlaps $R$ and $q$ at the optimal 
objective function
\begin{equation}
q(1-1/\alpha) - \frac{R^2}{q^{\star}} (1+\frac{1}{\alpha q^{\star}}) = 0 \ .
\end{equation}
Integrating \eqref{eq:Mprime} now gives up to a constant
\begin{equation}
\label{eq:M}
{\mathcal M}_v[\rho] (x)=-\frac{1}{\alpha v} \left( \frac{R}{q^{\star}} x -\frac{1}{2} (x-R)^2  \right) \ ,
\end{equation}
from which the optimal objective function can be obtained easily based on the relation that  for a convex function $f(y)$, ${\mathcal M}[f]_v (x)=g(x)$ implies $f(y)=-{\mathcal M}[-g]_v (y)$. Inverting~\eqref{eq:M} gives again up to a constant
\begin{equation}
\rho_{\text{opt}}(\h)= (\h-\frac{R(1+q^{\star})}{q^{\star}})^2 \ .
\end{equation}
The optimal value of $R$ is specified by extremizing the Lagrangian~\eqref{eq:lagrange1}, giving $q=q^{\star}$. Multiplying any objective function by a constant or adding a constant to it does not affect the reconstruction, so 
the optimal objective function can also be written as 
\begin{equation}
\label{eq:optimal_rho_appendix}
\rho_{\text{opt}}(\h) = \h^2 - 2 \delta \h \ ,
\end{equation}
with $\delta= (1+q^{\star}) \sqrt{\frac{q^{\star}(\alpha-1)}{\alpha q^{\star} +1}}$.

Of course $q^{\star}$ is not known when reconstructing the couplings. We set out considering $\delta$ in \eqref{eq:optimal_rho_appendix} a free parameter, which needs to be determined.
The free energy for this particular objective function is 
\begin{align} - f = \mbox{extr}_{q,v,R} \left[\frac{q-R^2/q^{\star}}{2v} - \alpha \frac{R^2+q-2 R \delta - 2 \delta^2 v}{2v+1} \right]\ .  
\end{align} 
with saddle-point equations giving 
\begin{align}
v&=\frac{1}{2(\alpha-1)} \\
 R&=\frac{q^{\star}\delta}{1+q^{\star}}\\
q&=\frac{\alpha q^{\star}+1}{(1+q^{\star})^2(\alpha-1)} \delta^2 \ .  
\end{align} 
The last result is crucial as the overlap $q$ of reconstructed couplings can be computed without knowing the underlying couplings. $q^{\star}$ can be determined from a simple linear fit of $q$ against $\delta^2$, which determines the offset parameter in the optimal objective function \begin{equation} \label{eq:delta}
\delta= (1+q^{\star}) \sqrt{\frac{q^{\star}(1-\alpha)}{\alpha q^{\star} +1}} \ .  
\end{equation}

\begin{figure}[tbh!]  
\includegraphics[width = .4\textwidth]{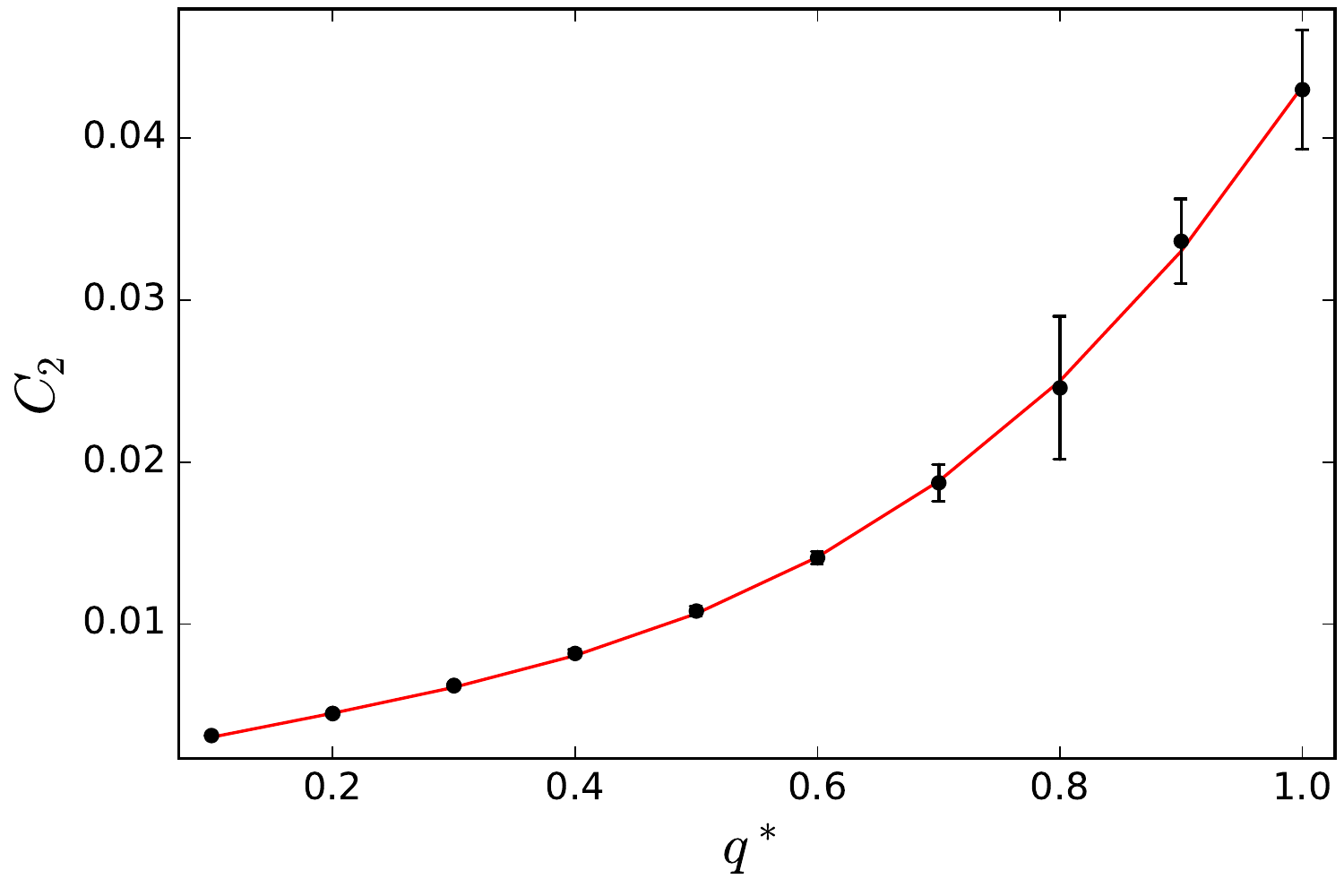} 
\caption{\textbf{Mean square spin-spin correlation.} We consider the squared spin-spin correlations averaged over all spin pairs defined by~\eqref{eq:ms_spincorr} for a system of $N=100$ spins from which $M=500$ samples are taken. The symbols indicate the mean and standard error over $10$ realisations of the couplings against the variance of the couplings $q^{\star}$. The line gives the result of a least-square fit to a 4th-order polynomial. Given the mean squared spin-spin correlation $C_2$ observed in a single realisation of the disorder, the value of $q^{\star}$ can be found by solving this polynomial for the observed value of $C_2$. The quality of this estimate will be worse at increasing values of $q^{\star}$ as the sample-to-sample fluctuations of  the squared spin-spin correlations increase.
 \label{fig_chi} } \end{figure}

There is a second way to determine the optimal value of $\delta$, which does not require optimizing the objective function \eqref{eq:optimal_rho_appendix} at different values of $\delta$.  If the variance of couplings $J_{ij}^{\star}$ of a spin glass model with Gaussian couplings (the Sherrington-Kirckpatrick model) were fixed to be $1$, $q^{\star}$ would be the square of the inverse temperature. The task is thus to determine the temperature parameter at which an observed set of $M$ spin configurations were taken, not knowing the concrete realisation of the underlying couplings but only the distribution from which the couplings were taken. We sketch a simple way to do this, based on the mean square spin-spin correlation 
\begin{equation}
 \label{eq:ms_spincorr} C_2=\frac{1}{N(N-1)} \sum_{i<j} [\frac{1}{M}\sum_{\mu} s_i^{\mu} s_j^{\mu} - m_i m_j]^2 
\end{equation} 
with $m_i=\frac{1}{M}\sum_{\mu}s_i^{\mu}$. This quantity is closely related to the spin-glass susceptibility, and can be computed 
from a set of spin samples. (The spin-glass susceptibility includes also a contribution from the diagonal term with $i=j$.)  
At a given value of $\alpha=M/N$, \eqref{eq:ms_spincorr} can be computed easily from spin-samples generated at different 
values of $q^{\star}$. Figure \ref{fig_chi} shows $C_2$, calculated numerically for $\alpha=5$, against $q^{\star}$, along with a fit to a 4th-order polynomial. 
Given a set of samples for which we want to reconstruct the couplings, one can evaluate $C_2$ for these samples and read off the 
corresponding value of $q^{\star}$ from Figure \ref{fig_chi} , or more specifically, solve the fitted polynomial for $q^{\star}$.  

We used this approach to determine $q^{\star}$ in a single realisation of the couplings with $N=100$ and $q^{\star}=0.25$ as in Figure~\ref{fig1} and~\ref{fig2}.  We produce $M=500$ samples, resulting in $C_2=0.00497$, for which the fit shown in Figure~\ref{fig_chi} gives an estimate for $q^{\star}$ of $0.235$. This yields a threshold of $\delta=0.812$, compared to the optimal value of $\delta$, which is approximately $0.833$. How small the resulting difference in the reconstruction error is can be read off from Figure~\ref{fig1}. This approach also works at even smaller values of $\alpha$. For $\alpha=1.1$ we observe in a single realisation of the couplings and a set of $110$ samples $C_2=0.01214$, for which the fit (recomputed at the new value of $\alpha$) gives $q^{\star}=0.254$ and the estimated threshold of approximately $\delta=0.176$, compared to the correct threshold of approximately $0.175$. The resulting reconstruction error is $\epsilon=1.353$, compared to the reconstruction error with the optimal value of $\delta$ of $0.132$. At $\alpha=1$ the threshold \eqref{eq:delta} reaches zero.

\subsection*{3. The distribution of couplings}

The statistics of reconstructed couplings can be read off from the free energy~\eqref{eq:big_partition3} using standard arguments, giving the average fraction of couplings exceeding a threshold $a$ as 
\begin{equation}
\int D_{q^{\star}} J^{\star} \int Dt \frac{\int_a^{\infty} dJ \exp\{-\frac{1}{2}(\hat{q}_0+\hat{q}_1)J^2+\sqrt{\hat{q}_0} t J+\hat{R} J^{\star} J  \} }{\int_{-\infty}^{\infty} dJ \exp\{-\frac{1}{2}(\hat{q}_0+\hat{q}_1)J^2+\sqrt{\hat{q}_0} t J+\hat{R} J^{\star} J \} } \ .
\end{equation}
For the low-temperature limit we use the scaling of the conjugate order parameters \eqref{eq:conjugate_ops}
\begin{align}
\hat{q}_1+\hat{q}_0&=\beta/v \\
\hat{q}_0&=\frac{\beta^2}{v^2} (q_0 -R^2/q^{\star}) \nonumber\\
\hat{R}&=\frac{\beta R}{v q^{\star}} \nonumber \ ,
\end{align}
which turns the integrals over $J$ into saddle-point integrals with saddle-point equation
\begin{equation}
J=\frac{R J^{\star}}{q^{\star}} + \sqrt{q-R^2/q^{\star}} t \ .
\end{equation}
Since $t$ follows a Gaussian distribution with mean zero and unit variance, this result means that the reconstructed coupling $J$ is on average 
$\frac{R J^{\star}}{q^{\star}}$ (specifying the bias) and has a variance $q-R^2/q^{\star}$. 

Similarly, the distribution of $\h^{\mu}$ can be calculated, in the low-temperature limit their statistics is that of 
\begin{equation}
\mbox{argmin}_{\h} \left[ \frac{(\h-R-\sqrt{q_0}t)^2}{2 v} + \rho(\h)\right] \equiv {\mathcal P}_v[\rho] (R+\sqrt{q}t) \ ,
\end{equation}
where $t$ is a univariate Gaussian with zero mean. $ {\mathcal P}_v[\rho](x)$ is called the proximal map~\cite{Parikh:2014}.

\subsection*{4. The optimal objective function in the presence of a regularizer} 

We consider a quadratic regularizer (an ${\ell_2}$-regularizer) added to the objective function, so the objective function becomes
\begin{equation}
\label{eq:suplm_objective}
\sum_{\mu} \rho(\h^{\mu}) + \frac{\gamma}{2} \sum_j J_{j}^2 \ .
\end{equation}
The motivation for this additional term is to penalize large couplings. Such a term can arise, for instance, from a (Gaussian) Bayesian prior.
The parameter $\gamma$ needs to be adjusted such that the self-overlap $q=\frac{1}{N}\sum_j  J_{j} J_{j}$ equals $q^{\star}$, which we assume to be known. 

This additional term only leads to a small alteration in the evaluation of the partition function~\eqref{eq:big_partition1}. 
The integrals over couplings in~\eqref{eq:gdef} become 
\begin{equation}
\label{eq:gS_reg}
e^{g_S(\{\hat{q}_{ab},\hat{R}_a\})} = \int D_{q^{\star}}J^{\star} \prod_a \int dJ^a \exp \{ -\frac{\gamma \beta}{2}\sum_a  (J^a)^2+ \sum_a i \sum_{a \leq b} \hat{q}_{ab} J^a J^b + i \sum_a \hat{R}_a J^a J^{\star}   \} 
\end{equation}
and the replica-symmetric free energy~\eqref{eq:big_partition4} becomes
\begin{align}
\label{eq:big_partition_reg}
-\beta f &\equiv  \frac{1}{N}\llangle \ln Z(\textbf{s}^1,\ldots,\textbf{s}^M) \rrangle \mbox{extr}_{q_1,\hat{q}_1,q_0,\hat{q}_0,R,\hat{R}} 
\left[ \frac{1}{2} q_1 \hat{q}_1 + \frac{1}{2} q_0 \hat{q}_0 -R \hat{R} \right. \nonumber\\
&\left.+ \frac{1}{2} \ln(2\pi) -  \frac{1}{2} \ln(\hat{q}_1+\hat{q}_0+\gamma \beta) 
+\frac{1}{2} \frac{\hat{q}_0+q^{\star} \hat{R}^2}{\hat{q}_1+\hat{q}_0+\gamma \beta} 
+\alpha \int Dt \ln \left[ \int \frac{d\h}{\sqrt{2\pi(q_1-q_0)}} \exp\{- \frac{(\h-R-\sqrt{q_0}t)^2}{2(q_1-q_0)}  -\beta \rho(\h)  \}
\right]
\right] \ ,
\end{align}
eliminating the conjugate order parameters $\hat{q}_0,\hat{q}_1,\hat{R}$ in the limit $\beta \to \infty$ as before lead to a free energy~\eqref{eq:free_energy}
\begin{align}
\label{eq:free_energy_reg}
- f &= \lim_{\beta \to \infty} \frac{1}{\beta}\llangle \ln Z(\textbf{s}^1,\ldots,\textbf{s}^M) \rrangle\\
&=\mbox{extr}_{q,v,R} \left[\frac{q-R^2/q^{\star}}{2v}  -\frac{\gamma}{2} q - \alpha \int Dt  {\mathcal M}_v[\rho](R+\sqrt{q}t) \right] \ .
\end{align}
The value of the regularization parameter $\gamma$ is determined such that the value of the self-overlap $q$ at the saddle-point equals $q^\star$. 
Other types of regularizing terms can be treated 
analogously. 

The evaluation of the optimal function $\rho(\h)$ in the presence of a regularizing term proceeds as in section 2. We note that jointly varying both the objective function \emph{and}  the regularisation parameter $\gamma$ introduces a trivial gauge degree of freedom, as multiplying both $\gamma$ and $\rho(\h)$ by some factor 
leaves the minimum of the objective function \eqref{eq:suplm_objective} unchanged. 
We find the same quadratic form as in the absence of the regularizing term (up to an arbitrary multiplicative constant set to unity, thus fixing the gauge)
\begin{equation}
\label{eq:optimal_rho_reg}
\rho_{\text{opt}}(\h) = \h^2 - 2 \delta \h \ ,
\end{equation}
but with an optimal value of the threshold $\delta= \frac{R(1+(1-\gamma v) q^{\star})}{(1-\gamma v) q^{\star}}$. 
The free energy for an arbitrary value of the threshold $\delta$ is
\begin{align} 
- f = \mbox{extr}_{q,v,R} \left[\frac{q-R^2/q^{\star}}{2v} -\frac{\gamma}{2}q- \alpha \frac{R^2+q-2 R \delta - 2 \delta^2 v}{2v+1} \right]
\end{align} 
with saddle-point equations
\begin{align}
\frac{1}{v} - \gamma - \frac{2 \alpha}{2v+1}&=0 \\
-\frac{R}{q^{\star} v}- \frac{2 \alpha(R-\delta)}{2 v+1}&=0\\
\frac{q-R^2/q^{\star}}{2 v^2}+ \frac{2\alpha \delta^2}{2v+1}+2\alpha\frac{R^2+q-2R\delta-2 v \delta^2}{(2v+1)^2} &=0
\end{align}
which are solved by
\begin{align}
v&=\frac{-(\gamma+2(\alpha-1))+\sqrt{(\gamma+2(\alpha-1))^2+8\gamma} }{4\gamma} \\
R&=\frac{(1-\gamma v) q^{\star}}{1+(1-\gamma v) q^{\star}} \delta \\
q&=\frac{R^2/q^{\star} +2 v (1-\gamma v) \delta^2 + \frac{1}{\alpha} (1-\gamma v)^2 (R^2-2 R\delta-2\delta^2 v)  }{1- (1-\gamma v)^2/\alpha}
\end{align}
Since the self-overlap $q=q^{\star}$ is fixed, the minimal reconstruction error coincides with the maximum overlap $R$. 
This maximum is reached for $\delta \to \infty$ and $\gamma \to \infty$ 
with $x \equiv \delta/\gamma$ remaining constant. Asymptotic analysis of the saddle point equations yields in this limit  
\begin{align}
v&=\frac{1}{\gamma+2\alpha}\\
R&=2 \alpha q^{\star} x\\
q&=4 \alpha (\alpha q^{\star} +1) x^2
\end{align}
and solving $q=q^{\star}$ gives $x=\sqrt{ \frac{q^{\star}}{ 4\alpha  (\alpha q^{\star} +1) } }$. The result for $R=\sqrt{\frac{\alpha q^{\star 3}}{ \alpha q^{\star} +1 }}$ is to be compared
to the result without the regularizing term, $R=\sqrt{\frac{(\alpha-1) q^{\star 3}}{ \alpha q^{\star} +1 }}$. 

For the numerical simulations in Fig~\ref{fig3} (bottom), each row of the matrix of couplings was computed individually. 
The regularisation parameter $\gamma$ was set for each row such that the self-overlap of the resulting couplings was equal to $q^{\star}$. For the optimal 
objective function, we set the offset $\delta$ to $20$, choosing larger values did not affect the results. 

\subsection*{5. Addendum: The high-temperature expansion revisited} 

Each term in high-temperature expansion \eqref{eq:hTexpansion} has been calculated exactly (to leading order in the thermodynamic limit). Nevertheless, this expansion remains problematic, as the different terms scale differently with $N$. Specifically, the second-order term is of order $N$, the fourth order term of order $N^2$, and so on. As a result, errors that do not contribute  in the thermodynamic limit to a high-order term can contribute to a low-order term. 

Inspired by the results of Bachschmid-Romano and Opper using the cavity method~\cite{bachschmid2017statistical}, we reformulate our expansion to address this problem. We aim to calculate the average 
\begin{equation}
\label{eq:cav_aim}
\sum_{\textbf{s}} \frac{1}{Z^{\star}} \exp\left\{ \frac{1}{\sqrt{N}}  \sum_{i<j}   J_{ij}^{\star} s_i s_j+  \frac{i}{\sqrt{N}} \sum_a \hat{\h}^{a} J_j^a  s_1 s_j \right\} \ ,
\end{equation}
where we have dropped the product over the pattern index $\mu$ for convenience. Rather than using replicas to represent the inverse of the partition function, 
we split the Hamiltonian into a part coupling to spin $1$ and a part not coupling to that spin, 
\begin{equation}
-\beta {\cal H} \equiv \frac{1}{\sqrt{N}}  \sum_{i<j}   J_{ij}^{\star} s_i s_j= 
\frac{1}{\sqrt{N}}  \sum_{j>1}   J_{1j}^{\star} s_1 s_j + 
 \frac{1}{\sqrt{N}}  \sum_{i>1,j>i}   J_{ij}^{\star} s_i s_j \ .
\end{equation}
The second term can be considered a `cavity Hamiltonian' ${\cal H}\woone \equiv  \frac{1}{\sqrt{N}}  \sum_{i>1,j>i}   J_{ij}^{\star} s_i s_j$, where the couplings 
to the first spin have been removed. We now perform the average with respect to the cavity Hamiltonian and expand the \emph{remaining} terms in a Taylor series. 

To this end, we also need to work out the difference between the partition function 
$Z^{\star} \equiv \sum_{\textbf{s}} \exp\{-\beta \H\}$ and the corresponding partition function of the cavity Hamiltonian, $Z^{\star}_{\backslash 1}$
\begin{align}
\frac{Z^{\star}}{Z^{\star}_{\backslash 1}} &= \frac{\sum_{\textbf{s}} \exp\{-\beta \H\} }{ \sum_{\textbf{s}} \exp\{-\beta \H\woone\} } = 
\frac{\sum_{\textbf{s}} \exp\{-\beta \H\woone\} \exp\{ \frac{1}{\sqrt{N}}  \sum_{j>1}   J_{1j}^{\star} s_1 s_j \}  }{ \sum_{\textbf{s}} \exp\{-\beta \H\woone\} } \nonumber\\
&=
\frac{1}{Z^{\star}\woone} \sum_{\textbf{s}} \exp\{-\beta \H\woone\} 
\left[
 1+\frac{1}{\sqrt{N}}  \sum_{j>1}   J_{1j}^{\star} s_1 s_j 
+ \frac{1}{2!} (\frac{1}{\sqrt{N}})^2  \sum_{j>1}  J_{1j}^{\star} s_j \sum_{k>1}  J_{1k}^{\star} s_k 
+\ldots 
\right] \nonumber\\
&=
1+\frac{1}{\sqrt{N}}  \sum_{j>1}   J_{1j}^{\star} \langle s_1 s_j \rangle\woone
+ \frac{1}{2!} (\frac{1}{\sqrt{N}})^2  \sum_{j>1,k>1}  J_{1j}^{\star} J_{1k}^{\star} \langle s_j s_k \rangle\woone
+\ldots \ .
\end{align}
The pointed brackets refer to averages with respect to the cavity Hamiltonian 
$\H_{\backslash 1}$. Odd-order terms are zero as $s_1$ does not couple to any other spins under this Hamiltonian. For even-order terms, we decompose four-spin averages like 
\begin{equation}
\langle s_j s_k s_l s_m \rangle\woone \approx \langle s_j s_k \rangle\woone \langle s_l s_m \rangle\woone + \langle s_j s_l \rangle\woone \langle s_k s_m \rangle\woone 
+\langle s_j s_m \rangle\woone \langle s_k s_l \rangle\woone\ ,
\end{equation}
which assumes that connected four-point correlations are small, and analogously for higher-order correlations. 
The number of such contractions for a term of order $2n$ is $(2n-1)(2n-3)\dots$ giving a combinatorial factor
$\frac{(2n-1)(2n-3)\dots}{(2n)!}=\frac{1}{2^nn!}$, the same combinatorial factor as in the expansion in section 1. 
Resumming the Taylor series then gives 
\begin{equation}
\frac{Z^{\star}}{Z^{\star}\woone} = \exp\left\{ \frac{1}{2N} \sum_{j>1,k>1}  J_{1j}^{\star} J_{1k}^{\star} \langle s_j s_k \rangle\woone  \right\}
\end{equation}
Crucially, the different powers of this expansion are all of the same order. 

We now use the same expansion to compute~\eqref{eq:cav_aim}, which we rewrite 
(with a shorthand $G_j \equiv  J_{1j}^{\star} + i \sum_a \hat{\h}^a J^a_j$)
\begin{align}
\label{eq:cav_result}
&\sum_{\textbf{s}} \frac{Z^{\star}\woone}{Z^{\star}} \frac{1}{Z^{\star}\woone} e^{-\beta \H\woone}
\exp\left\{\frac{1}{\sqrt{N}} \sum_{j>1} G_j s_1 s_j \right\} 
=\exp\left\{-\frac{1}{N} \sum_{j>1,k>1}  J_{1j}^{\star} J_{1k}^{\star} \langle s_j s_k \rangle\woone 
+\sum_{j>1,k>1}  G_j G_k \langle s_j s_k \rangle\woone\right\} \nonumber\\
&=\exp\left\{
\frac{i}{N} \sum_{j>1,k>1}  J_{1j}^{\star} \sum_a \hat{\h}^a J^a_k \langle s_j s_k \rangle\woone
-\frac{1}{2N} \sum_{a,b} \hat{\h}^a \hat{\h}^b \sum_{j>1,k>1}  J_{j}^a J_{k}^b \langle s_j s_k \rangle\woone \right\}
\end{align}
The diagonal terms in these sums from $\langle s_j s_j \rangle\woone=1$ give back the 
previous result~\eqref{eq:s_average_exponent} and~\eqref{eq:patternaverage_alphag1}, 
however, with decreasing temperature (increasing $q^{\star}$) the off-diagonal terms 
also play a role. The calculation described in section 1. can be performed with the result 
\eqref{eq:cav_result} instead of~\eqref{eq:s_average_exponent} and~\eqref{eq:patternaverage_alphag1} along the lines of~\cite{bachschmid2017statistical}:
By an orthogonal transformation of the vector of couplings $ J_{j}$ the matrix of correlations $\langle s_j s_k \rangle\woone$ 
can be diagonalized, yielding (up to an additive constant) the free energy \eqref{eq:free_energy}. However, the resulting order parameters have a different physical interpretation. Denoting the 
order parameters emerging here with a tilde we have 
$\tilde{q}=\frac{1}{N}\sum_{j,k}\llangle  J_{j} J_{k} \langle s_j s_k \rangle\woone \rrangle$ instead of~\eqref{eq:qdef} and $\tilde{R}=\frac{1}{N}\sum_{j,k}\llangle  J_{j} J_{1k}^{\star} \langle s_j s_k \rangle\woone \rrangle$ instead of~\eqref{eq:Rdef}. The link between the self-overlap $q$ and the overlap $R$ and these order parameters 
turns out to be simple (at least for fully connected models and without a regularizing term). Introducing suitable source terms into the partition function (see~\cite{bachschmid2017statistical}) 
gives $\tilde{R}=R$ and $\tilde{q}=q+(q q^{\star}-R^2)$. 
\end{document}